\documentclass[a4paper,man,floatsintext,longtable,noextraspace,10pt]{apa6}

\usepackage[english]{babel}
\usepackage[utf8x]{inputenc}
\usepackage{amsmath}
\usepackage{graphicx}
\usepackage[colorinlistoftodos]{todonotes}
\usepackage{hyperref}

\usepackage{booktabs}
\usepackage{longtable}
\usepackage{array}
\usepackage{multirow}
\usepackage{wrapfig}
\usepackage{float}
\usepackage{colortbl}
\usepackage{pdflscape}
\usepackage{tabu}
\usepackage{threeparttable}
\usepackage{threeparttablex}
\usepackage[normalem]{ulem}
\usepackage{makecell}
\usepackage{xcolor}
\NewDocumentCommand\citeproctext{}{}

\makeatletter
\let\@cite@ofmt\@firstofone
\def\@biblabel#1{}
\def\@cite#1#2{{#1\if@tempswa , #2\fi}}
\makeatother
\newlength{\cslhangindent}
\newlength{\csllabelwidth}
\newenvironment{CSLReferences}[2] % #1 hanging-indent, #2 entry-spacing
{\begin{list}{}{%
  \setlength{\itemindent}{0pt}
  \setlength{\leftmargin}{0pt}
  \setlength{\parsep}{0pt}
  \ifodd #1
  \setlength{\leftmargin}{\cslhangindent}
  \setlength{\itemindent}{-1\cslhangindent}
  \fi
  \setlength{\itemsep}{#2\baselineskip}}}
{\end{list}}
\usepackage{calc}

\shorttitle{}

\usepackage{tikz}
\usetikzlibrary{shapes.geometric, arrows, positioning, calc, shapes.multipart}

\tikzstyle{datasource} = [rectangle, rounded corners, minimum width=2.5cm, minimum height=1cm,text centered, draw=blue!60, fill=blue!5]
\tikzstyle{process} = [rectangle, minimum width=3cm, minimum height=1cm, text centered, draw=green!60, fill=green!5]
\tikzstyle{decision} = [diamond, minimum width=3cm, minimum height=1cm, text centered, draw=red!60, fill=red!5]
\tikzstyle{arrow} = [thick,->,>=stealth]
\tikzstyle{output} = [rectangle, rounded corners, minimum width=3cm, minimum height=1cm, text centered, draw=purple!60, fill=purple!5]

\begin{document}
\thispagestyle{otherpage}

%\maketitle

% Change \title, \date, remove \author:
\begin{large}
\textbf{Estimating transformative agreement impact on hybrid open access: A comparative large-scale study using Scopus, Web of Science and open metadata}
\end{large}

% Add below \maketitle:
\newcommand{\orcid}{%
  \begingroup\normalfont
  \includegraphics[height=0.9em]{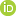}% removed .png extension
  \endgroup
}

Najko Jahn\textsuperscript{1}\textsuperscript{*} 
(\orcid{} \href{https://orcid.org/0000-0001-5105-1463}{\color{black}{0000-0001-5105-1463}})

\textsuperscript{1} Göttingen State and University Library, University of Göttingen, Germany. \\

\textsuperscript{*} Correspondence: \href{mailto:najko.jahn@sub.uni-goettingen.de}{\color{black}{najko.jahn@sub.uni-goettingen.de}} 

\section*{Abstract}
{This study compares open metadata from hoaddata, an openly available dataset based on Crossref, OpenAlex and the cOAlition S Journal Checker Tool, with proprietary bibliometric databases Scopus and Web of Science to estimate the impact of transformative agreements on hybrid open access publishing. Analysing over 13,000 hybrid journals between 2019-2023, the research found substantial growth in open access due to these agreements, although most articles remain paywalled. The results were consistent across all three data sources, showing strong correlations in country-level metrics despite differences in journal coverage and metadata availability. By 2023, transformative agreements enabled the majority of open access in hybrid journals, with particularly high adoption in European countries. The analysis revealed strong alignment between first and corresponding authorship when measuring agreement uptake by publisher and country. This comparative approach supports the use of open metadata for large-scale hybrid open access studies, while using multiple data sources together provides a more robust understanding of hybrid open access adoption than any single database can offer, overcoming individual limitations in coverage and metadata quality.}

{\textbf{Keywords}: bibliometric data sources, open metadata, hybrid open access, transformative agreements, curative bibliometrics}

\newpage

% QSS wants numbered sections
\setcounter{secnumdepth}{2}

\section{Introduction}\label{introduction}

Transformative agreements are a much-discussed library licensing model
for transitioning subscription-based journal publishing to full open
access. Although these agreements can vary considerably, they are mainly
aimed at hybrid journal bundles, enabling authors from participating
institutions to publish open access in these journals while providing
reading access to the entire portfolio (Borrego et al., 2021;
Hinchliffe, 2019). Following their initial proposal in 2015 (Schimmer et
al., 2015), the number of transformative agreements has grown
substantially, resulting in increased open access publishing in hybrid
journals (Jahn, 2025; McCabe \& Mueller-Langer, 2024; Rothfritz et al.,
2024). Prior to these agreements, only a few articles in hybrid journals
were openly accessible (Björk, 2012; Laakso \& Björk, 2016; Piwowar et
al., 2018), except for some European countries that had dedicated
funding policies and the SCOAP\(^3\) consortium for high-energy physics
journals (C.-K. (Karl). Huang et al., 2020; Kohls \& Mele, 2018;
Pinfield, 2015; Robinson-Garcia et al., 2020). Coordination efforts
through initiatives such as OA2020 and Efficiency and Standards for
Article Charges (ESAC) have further streamlined workflows between
library consortia and publishers (Campbell et al., 2022; Geschuhn \&
Stone, 2017). By April 2025, the ESAC Transformative Agreement
Registry\footnote{\url{https://esac-initiative.org/about/transformative-agreements/agreement-registry/}},
the primary community-driven evidence source, listed around 1,300
agreements between libraries and all the major commercial publishers and
leading society publishers(Rothfritz et al., 2024).

However, critics have raised concerns about these agreements,
particularly regarding perpetuating issues related to hybrid open access
such as charging twice for reading and publishing (``double dipping'')
(Asai, 2023), market concentration (Butler et al., 2023; Shu \&
Larivière, 2023), and the failure of most hybrid journals to convert to
full open access (Kiley, 2024; Momeni et al., 2021). Additional concerns
include reduced competition and incentives (McCabe \& Mueller-Langer,
2024; Schmal, 2024), and the continuation of publication fees that widen
gaps between well-resourced and under-resourced institutions (Babini et
al., 2022; Ross-Hellauer et al., 2022).

Consortia evaluations generally confirm increased open access through
transformative agreements and better coordination. However, they reached
mixed conclusions about the effectiveness. For instance, the British
Jisc evaluation (Brayman et al., 2024) found that the consortium's
transformative agreements had a significant impact on national open
access growth but a limited effect on the global open access transition,
suggesting a need to reevaluate its strategy. Similar policy
reconsiderations emerged in Norway and Sweden (Holden et al., 2023;
Widding, 2024). Meanwhile, the funder initiative cOAlition S recommended
ending financial support for these agreements in 2024, but still
considers hybrid open access to be compliant with funders' open access
policies. By contrast, the German DEAL consortium extended agreements
with major publishers until 2028, and countries in the Global South have
adopted transformative agreements (Muñoz-Vélez et al., 2024). Responding
to criticism, the Max Planck Digital Library recently outlined
strategies to evolve the original model, which continue to focus on
achieving fully open access while controlling costs and addressing
identified shortcomings (Dér, 2025).

These discussions highlight the need for robust data to evaluate
transformative agreements. Importantly, these agreements themselves can
help improve this evidence base, as researchers and libraries advocate
for including open metadata in negotiations to help avoid data analytics
from becoming a commodity controlled by publishers (Aspesi \& Brand,
2020; McCabe \& Mueller-Langer, 2024). For instance, the ESAC Workflow
Recommendations for Transformative and Open Access Agreements promote
using Crossref to share metadata that supports workflows including open
access licenses and funding information to support discovery and
monitoring (Geschuhn \& Stone, 2017). Simultaneously, the bibliometric
community pushes for more open metadata through this publisher-driven
DOI registration platform and continuously monitors progress (van Eck \&
Waltman, 2024).

This push for comprehensive metadata has yielded results. A growing
number of open scholarly data services that build on publisher-provided
metadata through Crossref have benefited from increasing metadata
coverage, although differences between publishers can be observed (van
Eck \& Waltman, 2024). OpenAlex stands out as a prominent example that
primarily retrieves updated records from Crossref and has gained
attention for its inclusivity, openness and comprehensive coverage
(Priem et al., 2022). Comparative studies confirm it as a suitable
alternative to proprietary bibliometric databases (Alperin et al., 2024;
Culbert et al., 2025). Notably, Céspedes et al. (2025) synthesised
evidence on both OpenAlex's strengths and its shortcomings for
bibliometric analysis. These include challenges with open access status
identification (Jahn et al., 2023; Simard et al., 2025), author
attribution (Culbert et al., 2025), institutional affiliations (Zhang et
al., 2024), and document type classification (Haupka et al., 2024).

While promising for general bibliometric research, measuring hybrid open
accesss and transformative agreements presents particular challenges.
With invoicing data largely unavailable (Kramer, 2024), previous studies
have relied on publication fee pricing lists or expenditures, and first
and/or corresponding author affiliations as proxies to attribute funding
to institutions (Butler et al., 2023; Haustein et al., 2024; Zhang et
al., 2022). Others manually obtained journals and institutions from
transformative agreement documents before retrieving eligible
publications from the Web of Science (Bakker et al., 2024). Recent work
has extended this approach and programmatically attributed open access
articles to transformative agreements by using data from the cOAlition S
Journal Checker Tool, which links journals and participating
institutions to transformative agreements included in the ESAC registry,
and the open metadata sources Crossref and OpenAlex (de Jonge et al.,
2025; Jahn, 2025). Yet, no systematic comparison exists of how open
versus proprietary bibliometric data sources perform when analysing the
same hybrid journals covered by transformative agreements to identify
comparative strengths and shortcomings of data sources.

The present study addresses this gap by comparing open and proprietary
data sources for evaluating transformative agreements. It adapts the
approach from Jahn (2025), which combines data from the cOAlition S
Journal Checker tool, Crossref as publication metadata source including
open access and OpenAlex as author affiliation data source, and applies
the same methods to the established databases Scopus and Web of Science.

This comparative approach examines different bibliometric data sources
through two primary research questions:

\begin{enumerate}
\def\labelenumi{\arabic{enumi}.}
\item
  How does the coverage of hybrid journals included in transformative
  agreements compare across Crossref, Scopus and Web of Science in terms
  of journal indexing and article representation?
\item
  To what extent do estimates of articles enabled by transformative
  agreements between 2019 and 2023 differ when using different
  affiliation data sources (OpenAlex, Scopus and Web of Science) and
  different authorship attribution methods (first versus corresponding
  authorship)?
\end{enumerate}

By analysing more than 13,000 hybrid journals across these sources, this
study provides insights into whether consistent results can be obtained
when measuring open access adoption across publishers and countries. The
findings enhance understanding of transformative agreements' impact on
the transition to full open access while revealing the comparative
strengths and limitations of different bibliometric databases in
measuring scholarly publishing's shift toward openness.

\section{Data and methods}\label{data-and-methods}

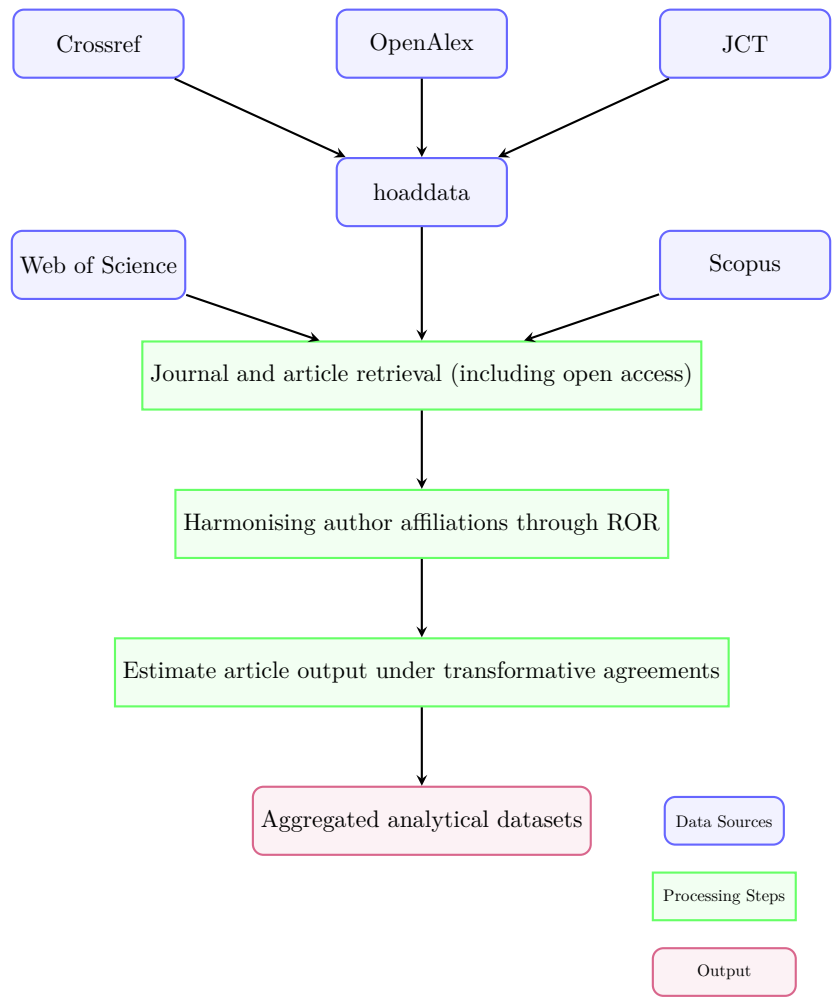
\begin{figure}[htbp]
\centering
\begin{tikzpicture}[node distance=2cm, thick,scale=1, every node/.style={scale=0.9}]    
% Data Sources
    \node (crossref) [datasource] {Crossref};
    \node (openAlex) [datasource, right=of crossref] {OpenAlex};
    \node (jct) [datasource, right=of openAlex] {JCT};
    \node (wos) [datasource, below=of crossref] {Web of Science};
    \node (scopus) [datasource, below=of jct] {Scopus};
    
    % Processing Steps
    \node (hoaddata) [datasource, below=1.5cm of $(crossref)!1!(openAlex)$] {hoaddata};
    \node (kb) [process, below=1cm of $(wos)!0.5!(scopus)$] {Journal and article retrieval (including open access)};
    
 % Matching Process - Fixed line break syntax
    \node (matching) [process, below=1.5cm of $(hoaddata)!1!(kb)$] {Harmonising author affiliations through ROR};

 % Aggregation
    \node (transmatch) [process, below=1.5cm of $(matching)!1!(matching)$] {Estimate article output under transformative agreements};
        
    % Output
    \node (analysis) [output, below=1.5cm of $(transmatch)!1!(transmatch)$] {Aggregated analytical datasets};
    
    % Arrows
    \draw [arrow] (crossref) -- (hoaddata);
    \draw [arrow] (openAlex) -- (hoaddata);
    \draw [arrow] (jct) -- (hoaddata);
    \draw [arrow] (hoaddata) -- (kb);
    \draw [arrow] (wos) -- (kb);
    \draw [arrow] (scopus) -- (kb);
    \draw [arrow] (kb) -- (matching);
    \draw [arrow] (matching) -- (transmatch);
    \draw [arrow] (transmatch) -- (analysis);

    % Legend
    \node [datasource, scale=0.7] at ($(analysis)+(4,0)$) {Data Sources};
    \node [process, scale=0.7] at ($(analysis)+(4,-1)$) {Processing Steps};
    \node [output, scale=0.7] at ($(analysis)+(4,-2)$) {Output};
\end{tikzpicture}
\caption{Data processing workflow for comparing hybrid open access uptake across bibliometric data sources. The workflow shows how data from different sources (hoaddata, derived from Crossref, OpenAlex, Transformative Agreement Data dump used by the cOAlition S Journal Checker Tool (JCT), the Web of Science, and Scopus) were processed to enable comparative analysis.}
\label{fig:workflow}
\end{figure}

As shown in Figure \ref{fig:workflow}, the methodology involved
comparing hoaddata, an openly available collection of open research
information on hybrid open access, with the bibliometric databases
Scopus and Web of Science. This section introduces the initial data
sources, followed by a presentation of the necessary data processing
steps to obtain eligible articles enabled by transformative agreements
using author roles (first and corresponding) and harmonised affiliation
data.

\subsection{Data sources}\label{data-sources}

\subsubsection{hoaddata}\label{hoaddata}

hoaddata\footnote{\url{https://github.com/subugoe/hoaddata}}, developed
and maintained by the author to support open access monitoring and
research (Achterberg \& Jahn, 2023; Jahn, 2025), is an R data package
that regularly collects information on hybrid open access uptake from
multiple openly available data sources (Jahn, 2024). It combines
article-level metadata from Crossref (Hendricks et al., 2020) and
affilation metadata from OpenAlex (Priem et al., 2022) with
transformative agreement information from the Transformative Agreement
Data dump used by the cOAlition S Journal Checker Tool (JCT)\footnote{\url{https://journalcheckertool.org/transformative-agreements/}},
which links journal and institutional data about participating research
organisations to agreements in the ESAC Transformative Agreement
Registry\footnote{\url{https://esac-initiative.org/about/transformative-agreements/agreement-registry/}}.

hoaddata follows good practices for computational reproducibility using
R (Marwick et al., 2018). The package, which includes data, code, a test
suite and documentation, is openly available on GitHub. To ensure
computational reproducibility while aggregating the data, a GitHub
Actions continuous integration and delivery (CI/CD) workflow handles
data retrieval from the SUB Göttingen's open scholarly data warehouse
based on Google BigQuery, which provides high-performant programmatic
access to monthly snapshots of Crossref and OpenAlex.\footnote{\url{https://subugoe.github.io/scholcomm_analytics/data.html}}
The workflow has run regularly to fetch updates from these data sources
since 2022. The package version used in this study is 0.3, containing
data from the Crossref 2024-08 dump provided to Crossref Metadata Plus
subscribers and the OpenAlex 2024-08-29 monthly dump. It covers
agreements collected between July 2021 to July 2024 from the JCT. This
version including the computation log is available on GitHub
(https://github.com/subugoe/hoaddata/releases/tag/v.0.3).

\subsubsection{Web of Science}\label{web-of-science}

Clarivate Analytics' Web of Science (WoS) is a well-established
proprietary bibliometric database consisting of several collections
(Birkle et al., 2020). Web of Science is selective regarding journal
indexing with a focus on base research (Stahlschmidt \& Stephen, 2022;
Visser et al., 2021). The collections considered in this study were the
Science Citation Index Expanded (SCIE), Social Sciences Citation Index
(SSCI) and Arts \& Humanities Citation Index (AHCI).

These collections provide important data points for analysing open
access: author affiliations and roles, differentiation of journal
articles into document types representing different types of journal
contributions, such as original articles or reviews, and open access
status information derived from OurResearch's Unpaywall (Piwowar et al.,
2018), the same provider as OpenAlex. However, Web of Science lacks
information about journals and articles under transformative agreements.

For programmatic access to article-level data, this study used the
database of the Kompetenznetzwerk Bibliometrie (KB) in Germany. The KB
processes raw XML data provided by Clarivate Analytics, which are
ingested into an in-house PostgreSQL database under a uniform schema. To
support reproducibility, KB maintains annual snapshots of the database.
Accordingly, this study used the annual snapshot from April 2024
(wos\_b\_202404), which is considered to cover almost the entire
previous publication year (Schmidt et al., 2024).

\subsubsection{Scopus}\label{scopus}

Elsevier's Scopus, launched in 2004, is another widely used proprietary
bibliometric database for measuring research (Baas et al., 2020).
Similar to Web of Science, Scopus is selective, but its journal coverage
is much broader than that of the Web of Science collections considered
in this study, as it indexes a wider range of applied research journals
(Singh et al., 2021; Stahlschmidt \& Stephen, 2022; Visser et al.,
2021). With detailed metadata about article types, open access status
information derived from Unpaywall, author roles, and disambiguated
affiliations, Scopus also contains important data to assess open access
uptake, although no direct information regarding transformative
agreements was available at the time of the study.

This study used the Scopus annual snapshot from April 2024 as provided
by KB (scp\_b\_202404). The same KB curation effort as that for Web of
Science was applied to Scopus (Schmidt et al., 2024).

\subsection{Data processing steps}\label{data-processing-steps}

\subsubsection{Determining the hybrid journal publication
volume}\label{determining-the-hybrid-journal-publication-volume}

Following Jahn (2025), the starting point was a unified dataset of
several safeguarded JCT snapshots\footnote{\url{https://github.com/njahn82/jct_data}}.
JCT journal data were enriched with ISSN variants linked to ISSN-L
according to the ISSN agency. To identify hybrid journals, a
comprehensive exclusion of fully open access journals was performed
using multiple journal lists including the Directory of Open Access
Journals (DOAJ). The resulting hybrid journal data comprising more than
13,000 hybrid journals were made available via hoaddata and were used to
independently determine the publication volume for each database.

hoaddata relies on Crossref to obtain journal publication volume and
open access status through Creative Commons (CC) licence information
relative to the published version (``version of record''). The article
metadata included DOIs, publication dates, open access information, and
author roles and affiliations. Publication years were determined using
the earliest known date of publication in a journal. In hoaddata, this
corresponded to Crossref's issued date. For Web of Science and Scopus,
the earliest publication date was used where available, with Scopus
dates specifically determined by the KB through version tracking of the
raw data.

Many transformative agreements typically cover only certain types of
journal articles, in particular original research articles including
reviews (Borrego et al., 2021). Because of limited information on these
document types in open scholarly data (Haupka et al., 2024), hoaddata
used an extended version of Unpaywall's paratext recognition approach to
exclude non-scholarly content (Piwowar et al., 2018). To exclude
conference supplements, which are often not covered by transformative
agreements, only articles published in regular issues, indicated by
numerical pagination, were considered. For Web of Science and Scopus,
their established and mainly accurate document type classifications
(Donner, 2017; Maisano et al., 2025) were used to identify original
research articles and reviews, referred to as original articles
throughout this study.

\subsubsection{Identifying open access articles in hybrid
journals}\label{identifying-open-access-articles-in-hybrid-journals}

Articles in hybrid journals were considered open access when publishers
made them freely available under a CC license on their platforms. While
hoaddata obtained this information from Crossref license metadata, Web
of Science and Scopus relied on Unpaywall as evidence source. Unpaywall
also uses Crossref license metadata, but supplements them by parsing
publisher websites directly, addressing cases where publishers do not
provide machine-readable CC license information (Piwowar et al., 2018).
This additional parsing remains necessary despite transformative
agreement workflows recommending the deposition of CC licenses during
DOI registration (Geschuhn \& Stone, 2017). Both Web of
Science\footnote{\url{https://webofscience.help.clarivate.com/en-us/Content/open-access.html}}
and Scopus\footnote{\url{https://web.archive.org/web/20250426153752/https://blog.scopus.com/posts/scopus-filters-for-open-access-type-and-green-oa-full-text-access-option}}
defined hybrid open access consistently as content available under CC
licenses on publisher platforms, distinguishing it from bronze open
access that lack such explicit license information, or use
publisher-specific licenses (Piwowar et al., 2018).

\subsubsection{Harmonising author affiliations across
databases}\label{harmonising-author-affiliations-across-databases}

Author affiliations were retrieved for both the first and, if available,
the corresponding authors to prepare the linking between articles and
institutions covered by transformative agreements. To improve data
retrieval, JCT institution data were enriched with ROR-IDs from
associated institutions, such as university hospitals or institutes of
large research organisations such as the Max Planck Society, according
to OpenAlex's institution entity. Database-specific affiliation
identifiers were used to handle different address variants: ROR-IDs from
OpenAlex for hoaddata, affiliation enhanced names for Web of Science,
and Scopus Affiliation Identifier. Additionally, ISO country codes were
retrieved for each author's address to compile country-level statistics.

Because neither Web of Science nor Scopus supported ROR-ID at the time
of data retrieval, the institution identifier used by the JCT, a
two-step matching process was implemented to harmonise the affiliation
data. First, 2,782,540 articles from 6,457 institutions with ROR-IDs in
the JCT data since 2017 (according to hoaddata) were processed to map
the first authors' ROR-IDs to corresponding proprietary affiliation
identifiers in Web of Science and Scopus using DOI matching. Then, an
algorithm selected the most frequent ROR-ID and proprietary organisation
identifier pairs to handle multiple affiliations and organisational
hierarchy differences.

This process linked 6,375 ROR-IDs to 4,894 Scopus Affiliation IDs and
6,034 ROR-IDs to 2,422 enhanced affiliation strings in Web of Science.
Quality evaluation through random sampling of 50 pairs revealed an error
rate of 22\% for Web of Science (11 mismatches) and 6\% for Scopus
(three mismatches). Upon inspection, these mismatches primarily occurred
with less-represented institutions having only a few publications,
introduced through multiple affiliations of single authors. The
difference between databases suggests that Scopus's affiliation control
is more closely aligned with ROR than the Web of Science is with ROR.

\subsubsection{Estimating open access in hybrid journals covered by
transformative
agreements}\label{estimating-open-access-in-hybrid-journals-covered-by-transformative-agreements}

Based on these compiled matching tables, articles eligible under
transformative agreements could also be obtained from Web of Science and
Scopus, although they did not contain ROR-IDs used by the JCT. The
estimation of eligible articles followed Jahn (2025) and included
matching of both journals and participating institutions according to
the JCT. The matching also considered the duration of agreements
according to the ESAC registry, with only those matches where an
agreement was actually in place being considered for subsequent
analysis. A related study (de Jonge et al., 2025), applied to
publications funded by the Dutch Research Council (NWO) and validated
against internal invoicing data, confirmed that such matching can
accurately identify most articles under transformative agreements.

\subsection{Data records}\label{data-records}

As a result of the comprehensive data processing described above,
datasets on open access in hybrid journals included in transformative
agreements were aggregated for each database at the country and journal
levels by year. Table~\ref{tbl-methods_overview_tab} provides a general
overview of coverage between 2019 and 2023. The table categorises hybrid
journal coverage in terms of publishing activity. Journals with at least
one publication during the five-years period (``active journals'') were
further differentiated based on whether they published at least one
original research article or review (``active journals (original)''),
collectively marked as ``original article'' in the following. The table
also presents the number of hybrid journals that made at least one
original article open access (``active journals (original) with OA'').
These journals formed the basis for subsequent calculations of
article-level indicators.

\begin{table}

\caption{\label{tbl-methods_overview_tab}Coverage of hybrid journals in
transformative agreements 2019-23.}

\centering{

\centering
\begin{tabular}[t]{lrrr}
\toprule
\textbf{} & \textbf{hoaddata*} & \textbf{Web of Science} & \textbf{Scopus}\\
\midrule
\addlinespace[0.3em]
\multicolumn{4}{l}{\textbf{Hybrid journal coverage}}\\
\hspace{1em}Active journals & 12,890 & 8,655 & 11,888\\
\hspace{1em}Active journals (original) & 12,888 & 8,655 & 11,878\\
\hspace{1em}Active journals (original) with OA & 11,348 & 8,392 & 11,313\\
\addlinespace[0.3em]
\multicolumn{4}{l}{\textbf{Publication volume}}\\
\hspace{1em}Total published articles & 9,740,015 & 8,616,053 & 8,117,644\\
\hspace{1em}Original articles & 8,158,425 & 6,708,083 & 7,317,703\\
\addlinespace[0.3em]
\multicolumn{4}{l}{\textbf{Digital Object Identifier (DOI) coverage}}\\
\hspace{1em}Articles with DOI & 9,740,015 & 7,713,796 & 8,105,112\\
\hspace{1em}Original articles with DOI & 8,158,425 & 6,695,661 & 7,314,327\\
\addlinespace[0.3em]
\multicolumn{4}{l}{\textbf{Open Access (OA) metrics}}\\
\hspace{1em}OA articles & 998,699 & 1,112,758 & 974,099\\
\hspace{1em}Original OA articles & 969,817 & 1,019,784 & 922,578\\
\addlinespace[0.3em]
\multicolumn{4}{l}{\textbf{Original articles with affiliation data}}\\
\hspace{1em}First author articles & 7,242,542 & 6,294,855 & 7,232,017\\
\hspace{1em}Corresponding author articles & 5,534,207 & 6,291,441 & 6,898,487\\
\bottomrule
\multicolumn{4}{l}{\rule{0pt}{1em}\textsuperscript{*} Journal article metadata from Crossref, except affiliations from OpenAlex}\\
\end{tabular}

}

\end{table}%

While hoaddata only covered articles with a DOI, Scopus and Web of
Science publication indicators were calculated using their database
identifier. A subsequent comparison of DOI coverage shows that
non-original articles in Web of Science often lacked a DOI. This was
particularly the case for meeting abstracts, which are notably prevalent
in Health Sciences journals (Melero-Fuentes et al., 2025) and are not
indexed by Scopus (Donner, 2017). Open access indicators were aggregated
by DOI, because Unpaywall only collects information on open access
status for articles in Crossref. A closer look at original articles with
affiliation data reveals a lack of data on affiliations and author roles
in the case of OpenAlex, the affiliation data source used by hoaddata,
compared to Web of Science and Scopus. In particular, only approximately
two-thirds of the articles examined provided corresponding author
affiliations. The proportion for first authors was 89\%. At the time of
writing, OpenAlex disclosed limited coverage of corresponding authorship
data \footnote{\url{https://docs.openalex.org/api-entities/works/work-object/authorship-object\#is_corresponding}}.
Therefore, only first author data for hoaddata were considered in the
following analysis.

\subsection{Data analysis}\label{data-analysis}

Tidyverse tools (Wickham et al., 2019) for the R programming language (R
Core Team, 2024) were used throughout this largely automated data
collection and analysis process. Rank correlations were calculated using
the Hmisc package (Harrell Jr, 2003). Source code analysis, including
all queries used to obtain the data, is available on GitHub
(\url{https://github.com/njahn82/hoa_validation}). An interactive
supplement exploring correlations between the data sources examined by
country and publisher is available via HuggingFace Spaces:
\url{https://huggingface.co/spaces/najkoja/hoa_replication}.

\section{Results}\label{results}

This section first presents the indexing of hybrid open access by
comparing open metadata with the proprietary bibliometric databases
Scopus and Web of Science. Then, using the same methods, indicators at
the publisher and country levels are calculated independently for each
database and compared with each other.

\subsection{Coverage comparision}\label{coverage-comparision}

\subsubsection{Overview}\label{overview}

\begin{figure}[ht!]

\centering{

\includegraphics[width=0.99\linewidth,height=\textheight,keepaspectratio]{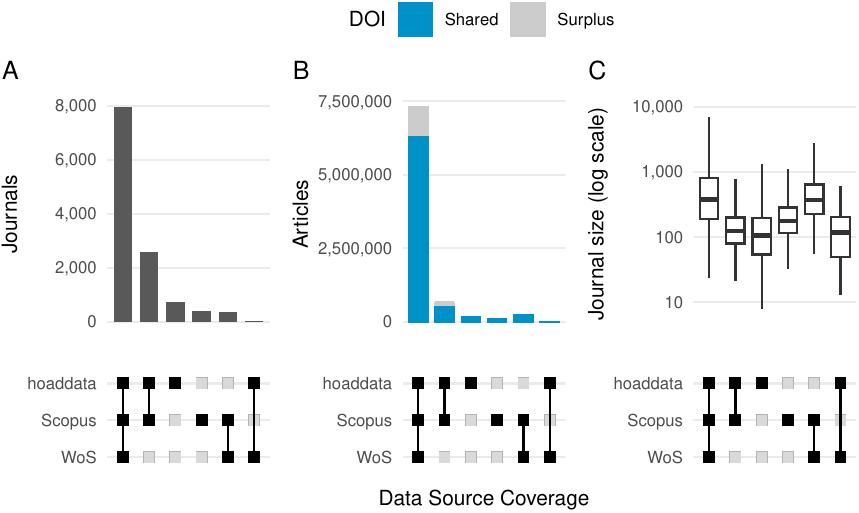}

}

\caption{\label{fig-upset_coverage_results}Comparison of hybrid journal
indexing by data source, 2019-2023. Only hybrid journals present in the
cOAlition S Transformative Agreement Data dump with at least one open
access article are considered. A) presents the number of journals, B)
the number of articles (DOI), distinguishing between shared DOI corpus
and surplus in hoaddata. Box plots (C) shows the five-years journal
article volume (log scale). Note that intersections sets with at least
30 journals are shown.}

\end{figure}

Figure \ref{fig-upset_coverage_results} presents the coverage of hybrid
journals included in transformative agreements, based on the
intersections of the examined data sources. The intersection sets of
journals (\ref{fig-upset_coverage_results}A) and articles
(\ref{fig-upset_coverage_results}B) are visualised as an UpSet graph
(Krassowski, 2020; Lex et al., 2014). The x-axis represents the set
intersections in a matrix layout. The analysis included hybrid journals
that published at least one open access article between 2019 and 2023,
based on open access status information from each database. Only
original articles were included in the analysis.

Journal coverage analysis revealed that 66\% (n = 7,970) of the hybrid
journals included in transformative agreements were indexed in all three
databases (Figure \ref{fig-upset_coverage_results}A). The second largest
set consisted of journals indexed in both hoaddata and Scopus,
comprising 21\% (n = 2,595) of hybrid journals. Notably, 6\% (n = 739)
of journals were exclusively contained in hoaddata, while another 6\% (n
= 748) were found only in Scopus. Of these, 354 were also available in
Web of Science. Upon inspection, the group of hybrid journals
exclusively covered in proprietary data sources mainly represented
hybrid journals for which no open access evidence could be retrieved
from Crossref, the open access evidence source for hoaddata.

In terms of article coverage, Figure \ref{fig-upset_coverage_results}B
shows the total publication volume per combination in terms of DOI
availability. The largest set of hybrid journals covering all three data
sources also represented the largest number of articles. In total, these
journals recorded 6,289,687 articles, represented by the blue bars. They
recorded 94\% of the original articles with a DOI indexed in Web of
Science, and 86\% in Scopus. Another 657,697 articles were exclusive to
the intersection between Scopus and hoaddata. Exclusively in hoaddata
were 177,110 articles, and exclusively in the proprietary databases were
325,194 articles.

Figure \ref{fig-upset_coverage_results}B also shows the surplus of
articles with DOI that were only available via hoaddata (grey area). In
the case of hybrid journals covered by all three data sources, 1,023,882
DOIs were present only in hoaddata. After validation at the DOI level
using KB databases and manual inspection, the main reason for missing
DOI coverage in the proprietary database was insufficient classification
of journal content as original articles during the compilation of
hoaddata. Particularly, letters and editorials could not be detected
fully. Moreover, paratext recognition failed for 37\% of DOIs to
identify non-scholarly content, such as front matters or reviewer lists,
which are generally not indexed by Scopus and Web of Science. To a
lesser extent, differences in publication and indexing dates were
reasons for non-overlapping DOIs.

Using overlapping DOIs, the publication volume between 2019 and 2023 was
calculated for each journal. Figure \ref{fig-upset_coverage_results}C
illustrates the distribution for each combination. It shows a large
spread of journal size across journals mutually covered by all three
data sources. These journals published more on average than journals
covered by only one or two data sources, with the exception of journals
exclusively covered by Scopus and Web of Science. In particular,
journals covered exclusively by hoaddata were substantially smaller.
Upon inspection, these were often newly launched hybrid journals, which
explains the relatively low five-year publication volume. An example is
\emph{Digital society} that published 86 articles. This hybrid journal
was launched in 2022, and had since been covered by various Springer
Nature transformative agreements.

\subsubsection{Coverage by publisher
portfolio}\label{coverage-by-publisher-portfolio}

\begin{figure}[ht!]

\centering{

\includegraphics[width=0.99\linewidth,height=\textheight,keepaspectratio]{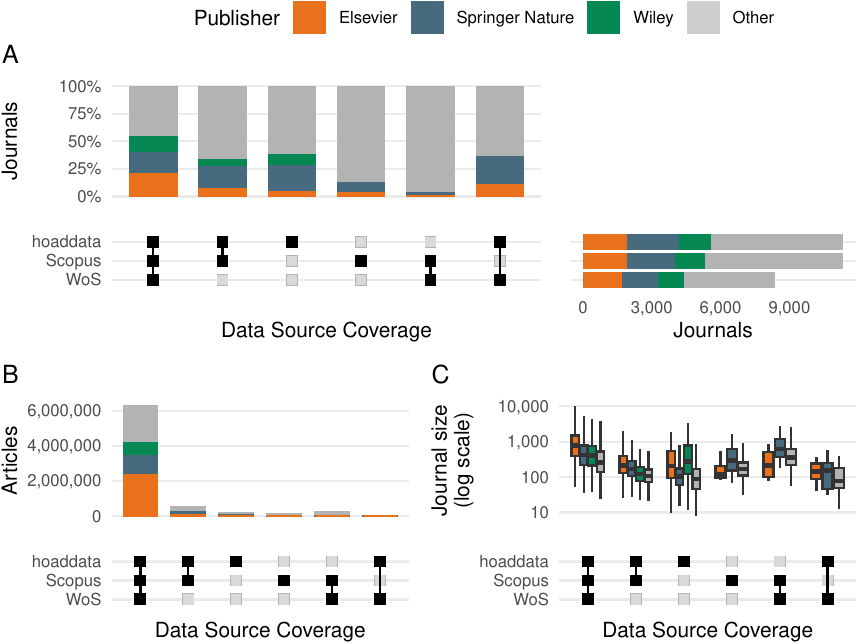}

}

\caption{\label{fig-upset_coverage_results_publisher}Comparison of
hybrid journal indexing by data source and publisher, 2019-2023. Only
hybrid journals present in the cOAlition S Transformative Agreement Data
dump with at least one open access article are considered. A) presents
the percentage of journals by publisher, B) the number of articles by
publisher (shared DOI). Box plots (C) shows the five-years journal
article volume (log scale) by publisher. Note that intersections sets
with at least 30 journals are shown.}

\end{figure}

Figure \ref{fig-upset_coverage_results_publisher} presents the coverage
of hybrid journals in transformative agreements across data sources from
2019 to 2023 with a focus on publisher portfolios. The analysis
highlights the dominance of the three largest publishers, Elsevier,
Springer Nature, and Wiley, which collectively accounted for 47\% of the
hybrid journals and 62\% of the articles published during this five-year
period. In terms of article volume, Elsevier led with 2,441,358 articles
(33\% of the total) published across 1,951 hybrid journals (16\% of the
total). Springer Nature followed with 1,247,578 articles (17\%) in 2,311
hybrid journals, although it recorded the largest number of hybrid
journals (19\%). Wiley accounted for 858,939 articles (12\%) in 1,382
hybrid journals (11\%). The remaining 54 publishers collectively
accounted for 2,749,847 articles (38\%) in 6,476 hybrid journals (53\%).

The three largest publishers, Elsevier, Springer Nature, and Wiley, were
best represented at the intersection of all three data sources
(hoaddata, Scopus, and Web of Science). Together, they comprised 4,384
hybrid journals (55\% of the intersectional set) and dominated article
coverage (n = 4,174,315; 66\%) as determined through shared DOIs. When
examining the publication volume per journal (Figure
\ref{fig-upset_coverage_results_publisher}C), Elsevier published, on
average, the largest journals.

A comparison of publisher portfolios across different indexing sets
demonstrates that publishers were not represented uniformly. Notably,
Springer Nature exhibited 519 hybrid journals exclusively indexed in
both hoaddata and Scopus. This set included journals from the Chinese
Academy of Science, German-language medical journals, and Eastern
European publications including the \emph{Journal of Mathematical
Sciences}, which also publishes English-language translations of
Russian-language works. Additionally, this subset included titles with a
broader disciplinary focus, such as \emph{SN Computer Science}, and
newly launched hybrid journals such as \emph{Nature Computational
Science}, which started in 2021 and were indexed in Scopus but not yet
in Web of Science. The set also captures discontinued journals,
providing further insights into the dynamics of journal publishing.

Examining publisher portfolios not covered by hoaddata, but present in
Scopus or Web of Science, identified several publishers with missing
Creative Commons (CC) license information in Crossref. In particular,
Emerald represented 322 journals with 86,409 articles, AIP Publishing
accounted for 24 journals with 64,898 articles, and World Scientific
recorded 87 journals with 42,531 articles. In total, 9 publishers did
not share CC licenses with Crossref and were therefore not represented
in hoaddata.

An inspection of individual journals also uncovered discrepancies in
Unpaywall's open access identification for certain publishers that
typically share CC license metadata with Crossref. Notably, some
subscription-only journals contained one or two articles erroneously
tagged as hybrid open access by Unpaywall, which was subsequently
reflected in Scopus and Web of Science. Examples of such
misclassifications include Elsevier's \emph{Journal of Bioscience and
Bioengineering}, and Springer Nature's \emph{Journal of Mechanical
Science and Technology}.

\subsection{Open Access Indicator
Comparison}\label{open-access-indicator-comparison}

This section examines the uptake of open access in hybrid journals,
focusing on the influence of transformative agreements across hoaddata,
Scopus, and Web of Science. The aim was to assess whether consistent
results could be derived from these data sources despite differences in
coverage and methodologies. Following Jahn (2025), indicators were
calculated for each data source and comprised the number and proportion
of open access articles, including those enabled by transformative
agreements, from 2019 to 2023. For Web of Science and Scopus, the impact
of transformative agreements was estimated using both first and
corresponding authorships, while hoaddata indicator calculations were
limited to first author affiliations.

\begin{figure}[ht!]

\centering{

\includegraphics[width=0.99\linewidth,height=\textheight,keepaspectratio]{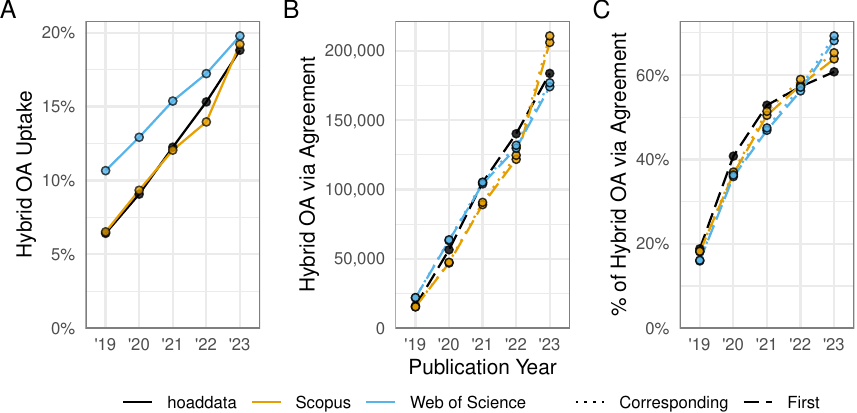}

}

\caption{\label{fig-uptake_overview}Development of open access in hybrid
journals in transformative agreements by data source and author role,
2019-2023. Figure shows the open access percentage (A), the number (B)
and the percentage (C) of open access articles enabled by transformative
agreements.}

\end{figure}%

\subsubsection{Overview}\label{overview-1}

Figure \ref{fig-uptake_overview}A shows a moderate growth of open access
in hybrid journals, which is consistent across hoaddata (black line),
Scopus (yellow line), and Web of Science (blue line). According to
hoaddata, hybrid open access uptake increased from 6.4\% (n = 85,071) in
2019 to 19\% (n = 302,358) in 2023. Similarly, Scopus recorded a growth
from 6.5\% (n = 84,648) in 2019 to 19\% (n = 322,850) in 2023. However,
Web of Science recorded higher open access uptake in the early years,
before converging to a similar level in 2023, from 11\% (n = 137,202) in
2019 to 20\% (n = 255,481) in 2023. This suggests that Web of Science
takes a different approach to labelling hybrid open access, which will
become clearer when results by publisher are presented later.

Figures \ref{fig-uptake_overview}B and C highlight that hybrid open
access by transformative agreements substantially increased between 2019
and 2023. Trends were consistent when measuring first (dashed line) and
corresponding authorship (dotted line). According to Scopus, 479,297
open access articles could be attributed to transformative agreements
based on first author metadata (increasing from 15,341 to 206,084) and
489,262 using corresponding author metadata (from 15,444 to 210,816).
Web of Science recorded 493,028 open access articles via transformative
agreements using first author metadata (increasing from 21,871 to
174,126) and 500,076 using corresponding author metadata (from 22,092 to
177,030). hoaddata, lacking corresponding author data, linked 501,649
articles to transformative agreements, increasing from 16,010 in 2019 to
183,757 in 2023.

Since 2021 (hoadata, Scopus) resp. 2022 (Web of Science), transformative
agreements have enabled the majority of open access articles in hybrid
journals. For the first authors, the share ranged between 61\%
(hoaddata), 64\% (Scopus), and 68 \% (Web of Science) by 2023. For the
corresponding authors, the shares were slightly larger, with Scopus
recording 65\% and Web of Science 69\% by 2023. However, substantial
hybrid open access was still facilitated outside transformative
agreements, likely through APCs paid from discretionary research funds
(Suber, 2012).

\subsubsection{Open access by
publishers}\label{open-access-by-publishers}

\begin{figure}[ht!]

\centering{

\includegraphics[width=0.99\linewidth,height=\textheight,keepaspectratio]{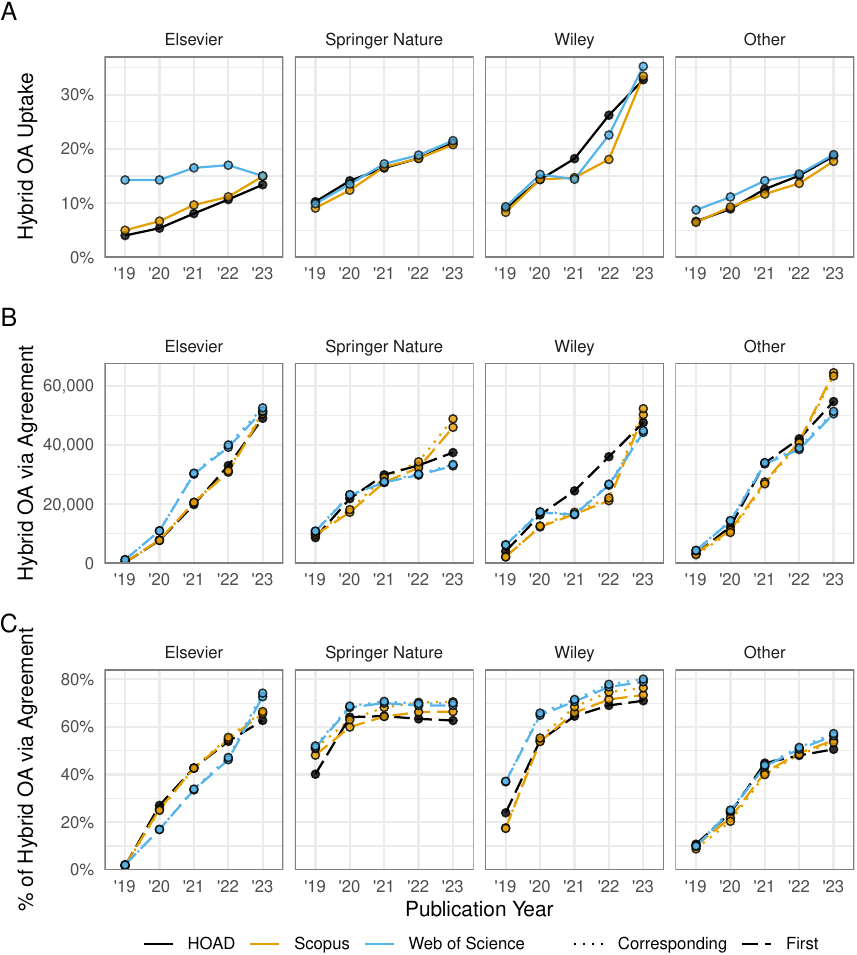}

}

\caption{\label{fig-uptake_publisher}Development of open access in
hybrid journals included in transformative agreements by data source,
author role and publisher, 2019-2023. Figure shows the open access
percentage (A), the number (B) and the percentage (C) of open access
articles enabled by transformative agreements.}

\end{figure}%

When considering open access trends by publisher (see Figure
\ref{fig-uptake_publisher}), the observed differences in early uptake
rates between hoaddata and Scopus compared to Web of Science can be
largely attributed to articles published in Elsevier hybrid journals,
the largest publisher in our sample. Both hoadata and Scopus reported a
steady increase in open access uptake between 2019 and 2023 (hoaddata
from 4\% to 13\%; Scopus from 5\% to 15\%). In contrast, Elsevier's
share remained relatively constant, increasing only slightly from 14\%
to 15\% according to Web of Science. Upon inspection, this discrepancy
was primarily due to articles under the publisher's open archive. These
articles, made freely available after an embargo period under Elsevier's
user license, were tagged as hybrid open access in Web of Science, even
though its documentation\footnote{\url{https://webofscience.help.clarivate.com/en-us/Content/open-access.html}}
specified that only articles under a CC license variant were considered.
Previous research (Haustein et al., 2024; Jahn et al., 2022) has shown
that Elsevier provided a substantial portion of its articles under this
license, explaining the relatively large and stable share of open access
over the years.

Differences in open access evidence are also apparent for Wiley.
Specifically, Web of Science and Scopus recorded a drop in 2021 and 2022
compared to hoaddata. For these two years, hoaddata reported 35,308 more
open access articles than Scopus and 32,491 more open access articles
than Web of Science. This discrepancy is presumably due to challenges in
fetching full-texts by Unpaywall, the open access evidence source for
Scopus and Web of Science. According to Unpaywall's software version
history, HTTP redirects from Wiley's publisher platform prevented
Unpaywall from parsing license information from full-texts.\footnote{See
  Unpaywall version history related to Wiley fixes:
  \url{https://github.com/search?q=repo\%3Aourresearch\%2Foadoi+wiley&type=commits}}
hoaddata, which relies solely on Crossref metadata for open access
identification, was unaffected by these issues.

Despite these differences in open access evidence, the three data
sources show consistent temporal trends in hybrid open access enabled by
transformative agreements (see Figure \ref{fig-uptake_publisher}B and
C). Wiley emerged as the fastest-growing publisher in terms of open
access uptake, with more than 30\% of articles in hybrid journals
reported as open access in 2023 across the examined data sources,
followed by Springer Nature. Elsevier recorded a later uptake,
consistent with the publisher's historical reluctance to engage in
negotiations with library consortia (Fraser et al., 2023). However, by
2023, the share of open access enabled by transformative agreements
appeared to stabilise for all three publishers (see Figure
\ref{fig-uptake_publisher}C). Interestingly, the differences between
first and corresponding authorships were more pronounced at the
publisher level. In Scopus, for example, the share of open access via
transformative agreements measured by the corresponding authorship was
higher for Springer Nature in 2023 than when using the first authorship.

\subsubsection{Open access by country}\label{open-access-by-country}

\begin{figure}[ht!]

\centering{

\includegraphics[width=0.99\linewidth,height=\textheight,keepaspectratio]{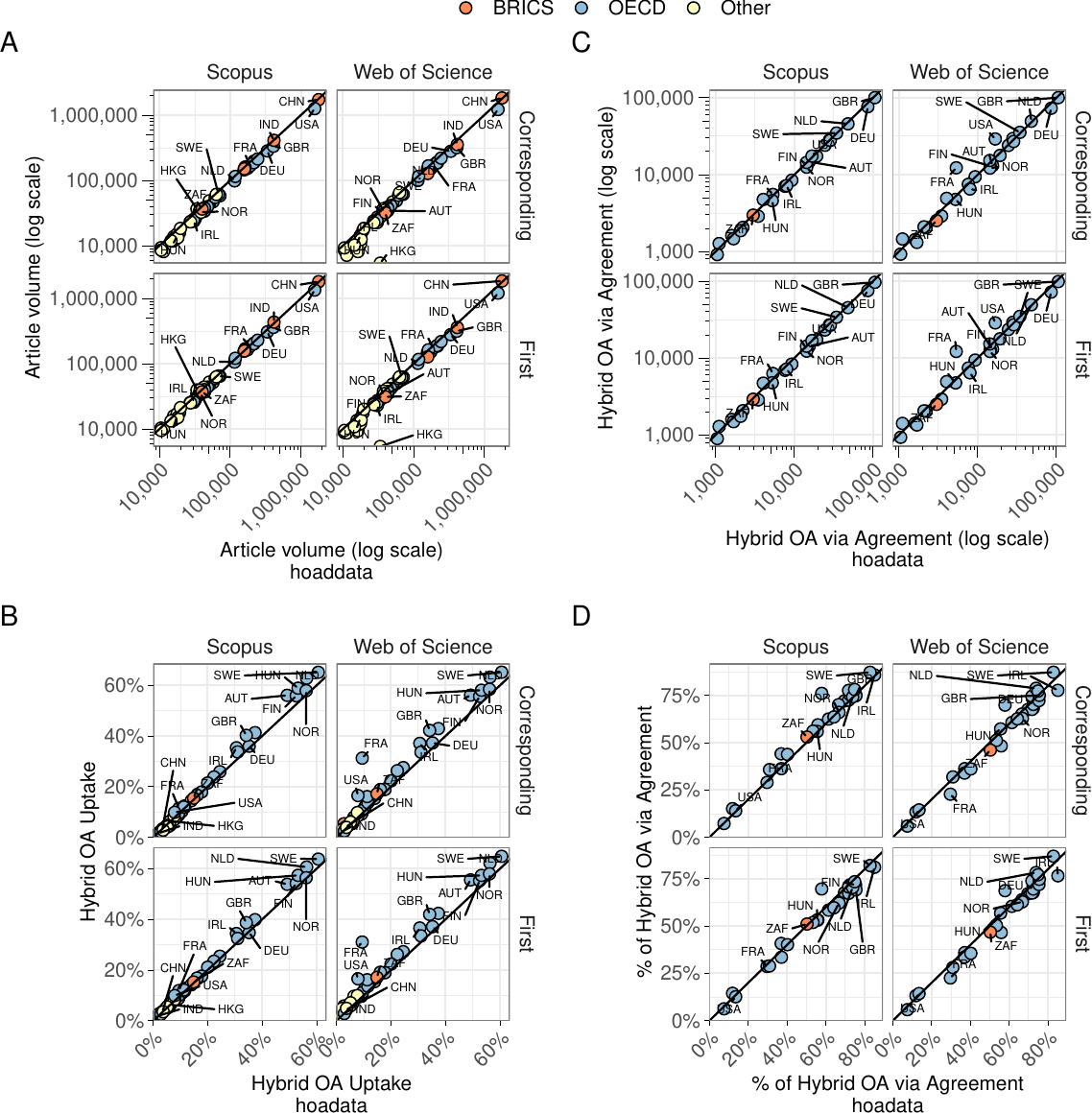}

}

\caption{\label{fig-uptake_country}Comparison of hybrid open access by
country, 2019-2023. Scatterplots distinguish between proprietary
databases Scopus and Web of Science, and author role. The x-axis shows
hoaddata indicators. A) Five-years article volume, B) open access
percentage in hybrid journals, (A and B limited to countries with
\textgreater{} 10.000 articles published), C) number and D) percentage
of open access articles enabled by transformative agreements (limited to
countries with \textgreater{} 1.000 open access articles enabled by
transformative agreements). Line represents line of equality. An
interactive version is accessible via:
\url{https://najkoja-hoa-replication.hf.space/}}

\end{figure}

When comparing countries, consistent patterns were observed across data
sources for the five-year period from 2019 to 2023. Figure
\ref{fig-uptake_country} presents hybrid open access indicators by
country, comparing hoaddata (x-axis) with Web of Science and Scopus
(y-axis). Indicators calculated from these proprietary databases are
shown for both the first and corresponding authors, with full counting
used to account for multiple country affiliations (Hottenrott et al.,
2021).

In terms of article output by country (see Figure
\ref{fig-uptake_country}A), strong positive correlations were observed
across data sources and author roles (Spearman rank correlation
\(\rho > .9\)). Between 2019 and 2023, China was the most productive
country, followed by the United States and, by a certain margin, India,
the United Kingdom, and Germany. The analysis of authorship roles
revealed minimal variation, indicating that the first and corresponding
author of an article were typically from the same country.

When examining the percentage of open access articles in hybrid journals
(see Figure \ref{fig-uptake_country}B), a different pattern emerged.
Authors affiliated with institutions from medium-sized European
countries, such as Sweden, the Netherlands, Finland, and Hungary, made a
large proportion of their articles open access. Germany and the United
Kingdom also had approximately 40\% of their output available as open
access. In contrast, non-OECD countries showed notably lower adoption of
hybrid open access, with South Africa being the only BRICS member (as of
2022) well-represented in the data. The United States also demonstrated
a relatively low proportion of open access articles. These findings were
consistent across all databases. However, France was better represented
in Web of Science, likely because of its agreement with Elsevier
starting in 2019, which allowed delayed open access under the
publisher's user license (Rabesandratana, 2019). This licence was not
classified as hybrid open access in either Scopus or hoaddata. In all
cases, Spearman rank correlations were \(\rho > .9\), indicating a high
level of correlation between the databases and authorship roles
considered.

Transformative agreements appeared to be a key driver of national open
access growth (see Figure \ref{fig-uptake_country}C and D). OECD members
accounted for the majority of open access articles enabled by
transformative agreements. As a notable exception, South Africa again
featured prominently, as the South African National Library and
Information Consortium (SANLiC) successfully negotiated transformative
agreements with major publishers from 2022 onward.\footnote{\url{https://sanlic.ac.za/read-and-publish-agreements/}}
The results were consistent across data sources. However, Wiley's open
access surplus in hoaddata led to better rankings for countries where
Wiley played a substantial role, such as Germany, where the DEAL
consortium negotiated its first transformative agreements with
publishers that started in July 2019.

Medium-sized European countries again showed a high proportion of hybrid
open access through transformative agreements (see Figure
\ref{fig-uptake_country}D), highlighting the impact of this licensing
model across all three data sources. In contrast, the United States had
a low proportion of hybrid open access enabled by agreements, suggesting
that a substantial number of open access articles from US-based authors
were likely financed through other means.

In all cases, strong positive correlations were observed using
Spearman's rank correlation: \(\rho > .9\) between data sources and
authorship roles, when considering countries with a minimum of 1,000
open access articles enabled by transformative agreements between 2019
and 2023. When this limitation was removed, the correlation remained
strong (\(\rho > .85\)). This difference may signal countries where only
a few institutions had transformative agreements in place, as opposed to
those participating in national consortia with broader participation.

\section{Discussion}\label{discussion}

This study of over 13,000 hybrid journals shows a substantial rise in
open access due to transformative agreements between 2019 and 2023,
although most articles remained paywalled. While transformative
agreements accounted for the majority of open access, many articles
continue to become open through the payment of individual publication
fees. Hybrid open access and transformative agreements remain
concentrated among a small group of large commercial publishers, with
European countries---alongside South Africa---showing high adoption
rates. In contrast, the three most productive countries, China, the
United States, and India, show a substantially lower adoption in
transformative agreements. Open questions remain as to whether this
uneven distribution reflects temporary implementation gaps, inherent
inequities in the transformative agreement model, or deliberate
avoidance of such agreements.

The findings were consistent across the investigated open data source
hoaddata, derived from Crossref and OpenAlex, and the established
proprietary bibliometric databases Scopus and Web of Science. This
aligns with previous studies (Akbaritabar et al., 2024; Alperin et al.,
2024) and rankings (van Eck et al., 2024). Overall, the results show
strong correlations by country affiliation, which supports the use open
metadata for large-scale analyses of hybrid open access. However, the
observed differences in journal coverage and metadata availability
warrant further discussion, affecting not only open data sources but
also proprietary databases when used in isolation.

The coverage analysis revealed that hybrid journals are well indexed in
all three data sources, particularly in terms of article coverage, which
reflects the dominance of major publishers whose established journal
portfolios are comprehensively indexed in proprietary databases (Bakker
et al., 2024; van Bellen et al., 2025). Differences emerge for journals
targeting practitioners or local non-English language communities, with
many such titles indexed exclusively in Crossref and Scopus. Using
Crossref as a bibliometric database in hoaddata demonstrated a
particular strength in identifying newly established hybrid journals, a
notable finding given that transformative agreements primarily target
existing subscription-based journals. The landscape of hybrid journal
publishing thus differs markedly from that of fully open access
journals. Comparing the coverage of OpenAlex, Scopus and Web of Science,
Simard et al. (2025) indicate that only half of the fully open access
journals listed in the DOAJ are also indexed in Scopus and Web of
Science. Notably, journals that charge no publication fees (``diamond
journals'') are absent from the selective Web of Science, which
reinforces existing disparities in the indexing of under-represented
research communities and regions in selective bibliometric databases
(Simard et al., 2025).

A frequently reported limitation of studying open access with less
selective databases is the lack of corresponding authorship information
(Fraser et al., 2023; Haucap et al., 2021; Shu \& Larivière, 2023).
However, this analysis demonstrates that indicators based on first
authors, which have often been used as a proxy for determining open
access funding, and corresponding authors show a high level of
correlation, reflecting disciplinary norms in scholarly publishing with
regard to contributions, author roles and positions (Larivière et al.,
2016; Zhang et al., 2022). The first author typically conduct the main
research underlying a paper, while the corresponding author often
supervises the research (Fox et al., 2018; Mattsson et al., 2010).
Unsurprisingly, measures based on first or corresponding authorships are
strongly correlated, suggesting that these authors share the same
country affiliation. In most cases, the first author is identical to the
corresponding author (Chinchilla-Rodríguez et al., 2024). Despite this
correlation, the study observed a recent slight increase in open access
articles attributed to transformative agreements by corresponding
authors over first authors across Springer Nature hybrid journals. This
prompts questions about how institutional open access sponsorship
practices influence author roles and assignments within co-author teams,
especially as funding opportunities vary (Gumpenberger et al., 2018).
Previous research has highlighted how institutionalised bibliometric
practices can affect the valuation of authorship positions (Helgesson,
2020), suggesting the need to monitor the potential influences of the
availability of open access funding on authorship roles (Maddi \& Silva,
2024).

Another critical data element in the study is affiliation data, which is
essential for estimating open access enabled by transformative
agreements. Although OpenAlex's affiliation coverage is less
comprehensive, which likely reduced the number of articles confidently
attributed to transformative agreements in this study, it still showed
high correlations with Scopus and Web of Science at the country level.
However, OpenAlex's native ROR-ID integration offers a distinct
advantage, allowing a more reliable identification of agreement-enabled
articles compared to Scopus and Web of Science, which require
reconciliation with proprietary organisation identifiers. Future studies
based on Web of Science will benefit from the recent integration of
ROR-IDs, announced by the end of 2024.

The database comparison revealed important discrepancies in open access
evidence. Crucially, not all publishers share CC licence metadata via
Crossref, a limitation that becomes apparent when contrasting Crossref
with Unpaywall's data in Scopus and Web of Science. While Unpaywall can
detect such gaps by parsing publishers' websites for open access
licences, it nevertheless missed a substantial number of CC-licensed
open access articles from Wiley journals indexed in Crossref, likely
because of parsing errors on the publisher's website. This resulted in
fewer open access articles being recorded in Scopus and Web of Science.
Further inconsistencies emerged between Scopus and Web of Science,
despite both relying on Unpaywall: Web of Science erroneously labels
Elsevier's delayed open access as hybrid, whereas Scopus correctly
categorises it. Notably, Scopus and hoaddata showed greater alignment
than that observed in a related comparison between Scopus and OpenAlex
(Alperin et al., 2024). This likely reflects the advantage of using of a
curated list to identify hybrid journals, rather than depending solely
on article-level open access tags.

The observed variations may compromise comparability not just in
research, but also across open access monitoring services, which have
grown over the years (Pampel \& Schneider, 2025; Salamoura \& Tsakonas,
2024). This study highlights, in line with previous research (Salamoura
\& Tsakonas, 2024), that design decisions underlying these services,
such as data sources used, update frequencies, and open access
definitions, are critical. Stakeholders should therefore advocate for
greater transparency from monitoring service providers in their
methodological choices and data processing decisions (Kramer, 2024).
Research and monitoring exercises might also benefit from avoiding
reliance on a single source. Instead, the selection of open access
evidence could be cross-verified using multiple sources and snapshots
that can be used to track changes in the data over time (C.-K. Huang et
al., 2020; Pölönen et al., 2020). Incorporating expert-curated journal
lists may also help reduce misclassification based on business models
(Visser et al., 2021).

It should be noted that the study's estimates of articles from
institutions involved in transformative agreements are approximations
because of a lack of access to invoice data, which is not usually shared
by library consortia and publishers (Kramer, 2024). Furthermore, the
open data sources for transformative agreements, including the cOAlition
S Journal Checker Tool and the ESAC registry, are based on voluntary
effort, crowd-sourced from various consortia. Additionally, while this
study observed strong correlations at the country level,
institutional-level correlations may differ due to matching inaccuracies
that can occur when working with long lists of participating
institutions and different disambiguation approaches for author
affiliations. Future research could explore institutional-level analysis
in more detail, though recent validation using Dutch research
information demonstrates the reliability of such an open approach for
assessing articles under transformative agreements (de Jonge et al.,
2025).

Despite these current limitations, the situation is likely to improve
with enhanced open metadata compliance attributable to evolving
standards and initiatives in support of negotiations with publishers,
particularly through the ESAC initiative and Barcelona Declaration on
Open Research Information, the situation is likely to improve. For
example, AIP Publishing and Emerald, whose hybrid journal portfolios
could not be included in this study due to a lack of CC licence
information, have recently begun sharing this metadata via Crossref.
Furthermore, during the 17th Berlin Open Access Conference (B17), an
international conference organised by the OA2020 Initiative, research
organisations and their library consortia emphasised the global
importance of open research information for achieving transparency. In
conclusion, this study has shown that open metadata are well-suited for
analysing the transition from subscription-based journal publishing to
full open access. But using them in conjunction with proprietary
databases provides a more robust understanding that overcomes individual
shortcomings in coverage and metadata quality.

\section*{Acknowledgement}\label{acknowledgement}
\addcontentsline{toc}{section}{Acknowledgement}

I would like to thank the two reviewers and the PREreview reviewers
(Rittman et al., 2025) for their valuable and helpful feedback on the
manuscript.

\section*{Author contributions}\label{author-contributions}
\addcontentsline{toc}{section}{Author contributions}

The author confirms the sole responsibility for the following:
Conceptualisation, Methodology, Formal analysis, Investigation, Data
Curation, Writing - Original Draft, Writing - Review \& Editing,
Visualization.

\section*{Competing interests}\label{competing-interests}
\addcontentsline{toc}{section}{Competing interests}

The author has participated in working groups of OA2020, ESAC and the
Barcelona Declaration as part of his professional responsibilities. No
funding was received from these initiatives for the present study, and
they had no influence on its design, data collection, analysis, or
interpretation of results.

\section*{Funding information}\label{funding-information}
\addcontentsline{toc}{section}{Funding information}

This work was supported by the Federal Ministry of Education and
Research of Germany (BMBF) under grants 16WIK2301E / 16WIK2101F. The
proprietary bibliometric data from Scopus and Web of Science were
provided by the German Competence Network for Bibliometrics under BMBF
grant 16WIK2101A.

\section*{Data and code availability}\label{data-and-code-availability}
\addcontentsline{toc}{section}{Data and code availability}

Data source code supplement is available via GitHub:
\url{https://github.com/njahn82/hoa_validation}. An interactive
dashboard, which allows the browsing of metrics by country and
publisher, is available on HuggingFace Spaces:
\url{https://huggingface.co/spaces/najkoja/hoa_replication}.

Please note that according to the licence terms for Scopus and Web of
Science, as provided by the German Competence Network for Bibliometrics,
only aggregated data accompanying the scientific publication can be made
publicly available for replication purposes.

Open metadata provided by Crossref, OpenAlex, and cOAlition S Journal
Checker Transformative Agreement Data are freely available for re-use.

\section*{References}\label{references}
\addcontentsline{toc}{section}{References}

\phantomsection\label{refs}

\end{document}